\documentclass[manuscript,screen]{acmart}

\AtBeginDocument{%
  \providecommand\BibTeX{{%
    \normalfont B\kern-0.5em{\scshape i\kern-0.25em b}\kern-0.8em\TeX}}}

\copyrightyear{2021} 
\acmYear{2021} 
\setcopyright{acmcopyright}\acmConference[Chinese CHI 2021]{The Ninth International Symposium of Chinese CHI}{October 16--17, 2021}{Online, Hong Kong}
\acmBooktitle{The Ninth International Symposium of Chinese CHI (Chinese CHI 2021), October 16--17, 2021, Online, Hong Kong}
\acmPrice{15.00}
\acmDOI{10.1145/3490355.3490358}
\acmISBN{978-1-4503-8695-1/21/10}

\begin{document}



\title[Think-Aloud Verbalizations and UX Problems: Effects of Language Proficiency]{Think-Aloud Verbalizations for Identifying User Experience Problems: Effects of Language Proficiency with Chinese Non-Native English Speakers}

\author{Mingming Fan}
\authornote{corresponding author}
\affiliation{%
  \institution{Hong Kong University of Science and Technology}
  \city{Hong Kong \& Guangzhou}
  \country{China}
}
\email{mingingfan@ust.hk}

\author{Lingyun Zhu}
\affiliation{%
  \institution{Rochester Institute of Technology}
  \city{Rochester, NY}
  \country{USA}
}
\email{lz2035@rit.edu}

\renewcommand{\shortauthors}{Fan and Zhu}

\begin{abstract}

Subtle patterns in users' think-aloud (TA) verbalizations (i.e., utterances) are shown to be telltale signs of user experience (UX) problems and used to build artificial intelligence (AI) models or AI-assisted tools to help UX evaluators identify UX problems automatically or semi-automatically. Despite the potential of such verbalization patterns, they were uncovered with native English speakers. As most people who speak English are non-native speakers, it is important to investigate whether similar patterns exist in non-native English speakers' TA verbalizations. As a first step to answer this question, we conducted think-aloud usability testing with Chinese non-native English speakers and native English speakers using three common TA protocols. We compared their verbalizations and UX problems that they encountered to understand the effects of language and TA protocols. Our findings show that both language groups had similar amounts and proportions of verbalization categories, encountered similar problems, and had similar verbalization patterns that indicate UX problems. Furthermore, TA protocols did not significantly affect the correlations between verbalizations and problems. Based on the findings, we present three design implications for UX practitioners and the design of AI-assisted analysis tools. 

\end{abstract}



\maketitle

\section{Introduction}

Think-aloud protocols (TAs) are widely used in usability testing to identify user experience (UX) problems~\cite{fan2020Survey,mcdonald2012exploring}. During TA usability testing, participants verbalize what they are thinking while working on the task with the test interface. Their verbalizations (i.e., utterances) provide access to their invisible thought processes, which are otherwise inaccessible to UX evaluators.
Despite the value of conducting TA usability testing, analyzing recorded TA sessions is often arduous and time-consuming~\cite{fan2020Survey,mcdonald2012exploring}. Traditional analysis methods are largely manual, which entail playing session recordings, listening to users' TA verbalizations, and observing other behavioral signals simultaneously to pinpoint UX problems. As it becomes increasingly easier to conduct a large amount of TA usability test sessions remotely via online platforms (e.g.,~\cite{usertesting_usertesting_2021,fullstory_fullstory_2021}), it is imperative to explore ways to improve traditional manual analysis methods. 

Toward this goal, researchers recently studied users' TA verbalizations (i.e., what users say) and their speech features (i.e., how they say it)~\cite{zhao2010keep,elling2012,cooke2010,hertzum2015thinking,Fan2021OlderAdults} and uncovered a series of subtle verbalization and speech patterns that are telltale signs of UX problems~\cite{fan2019concurrent,Fan2021OlderAdults}.
For example, when users encounter UX problems, they tend to verbalize utterances of their observations and remarks than other types of utterances (e.g., action description)~\cite{fan2019concurrent,Fan2021OlderAdults}.
Such TA verbalization patterns have been utilized to build artificial intelligence (AI) models to detect UX problems automatically~\cite{fan2020automatic}. Moreover, these patterns have also been leveraged to build human-AI collaborative analysis tools to better detect UX problems by combining the advantages of both AI and UX domain experts~\cite{fan2020vista}.

Despite the great potentials of TA verbalization patterns for automatically or semi-automatically detecting UX problems~\cite{fan2020vista,fan2020automatic}, these patterns were uncovered with \textit{native} English speakers~\cite{fan2019concurrent,Fan2021OlderAdults}. Compared to native English speakers (379 million)~\cite{Whatisth31:online, crystal2003english}, more people around the world speak English as a second language (753 million) (i.e., \textit{non-native} English speakers)~\cite{Whatisth31:online, crystal2003english}. 
Take the US as an example, almost half of the residents in America's largest cities speak a language other than English at home ~\cite{AlmostHa40:online}.
Over 151,000 employees from all over the world working in the US every year do not necessarily speak English as their first language~\cite{h1b:online, h2b:online}.
Consequently, it is not uncommon that non-native English speakers would participate in usability testing and be asked to think aloud in English.
Thus, it is important to understand whether verbalization patterns discovered among native English speakers~\cite{fan2019concurrent,Fan2021OlderAdults} still exist among non-native English speakers and whether there are any differences. 
Motivated by this problem, in this research, we took a first step to explore the following research question (RQ): 
\begin{itemize}
\item \textit{\textbf{RQ1}: How do English language groups (i.e., native and non-native speakers) affect TA verbalizations and UX problems?} 
\end{itemize}

Furthermore, three types of TA protocols are commonly used in usability testing~\cite{fan2019concurrent,mcdonald2012exploring,boren2000thinking}: 1) Ericsson and Simon's classic think-aloud protocol (CTA)~\cite{ericsson:1984}, which was used to uncover the subtle verbalization patterns indicative of UX problems among native English speakers~\cite{fan2019concurrent,Fan2021OlderAdults}, 2) the speech-communication protocol (SC)~\cite{boren2000thinking}, and 3) the interactive think-aloud protocol (ITA)~\cite{rubin2008handbook,dumas1999practical}. While participants are only reminded to ``keep talking'' in CTA, they receive speech tokens (e.g., ``Em hmm'‘’) from the moderator in SC or are constantly probed to answer questions (e.g., ``what are you looking for?'') from the moderator in ITA. In other words, participants experience different amounts of interventions while thinking aloud in 
these protocols. However, it remains unknown how TA protocols affect non-native and native English speakers' verbalizations and UX problems they experience. In this research, we took an initial step to explore: 
\begin{itemize}
\item \textit{\textbf{RQ2}: How do TA protocols (i.e., CTA, SC, and ITA) affect two English language groups' verbalizations and UX problems?} 
\end{itemize}

To answer two RQs, we conducted online think-aloud usability testing with 18 non-native and native English speakers. As non-native English speakers of different cultures might have different thinking and speaking behaviors, we focused on a subgroup of non-native English speakers---Chinese students who studied in US universities in this research as a first step to explore this problem space. 
Chinese students in US universities have taken a lion share of all international students enrolled in the past decades~\cite{Internat22:online} and speak English as a second language regularly in their study and daily life. 

During the study, participants of two language groups (i.e., non-native and native speakers) worked on tasks with three representative websites while thinking aloud using three TA protocols (i.e., CTA, SC, ITA) respectively. We transcribed their verbalizations (i.e., utterances), categorized them into five categories following prior studies~\cite{cooke2010,elling2012,fan2019concurrent,Fan2021OlderAdults}, identified UX problems that they encountered, and analyzed how different verbalization categories indicate UX problems.

Our results show that non-native English speakers' verbalizations were similar to those of native English speakers in terms of the relative proportions of different verbalization categories and the correlations between verbalization categories and UX problems. Furthermore, the trends between verbalization categories and UX problems were mostly consistent across three types of TA protocols. Based on the findings, we further discuss the implications for building AI models and human-AI collaborative UX data analysis tools. In sum, we make the following contributions:
\begin{itemize}
\item An initial understanding of how language groups affect TA verbalizations and their correlations with UX problems;
\item An initial understanding of how three types of TA protocols affect users' verbalizations and their correlations with UX problems. 
\end{itemize}

\section{Background and Related Work}
Our work was inspired and informed by related work in three areas: \textit{Types of Concurrent Think-Aloud Protocols}, \textit{Language Proficiency in Think-Aloud Studies}, and \textit{Users' Verbalizations in Think-Aloud Studies}.

\subsection{Three Types of Concurrent Think-Aloud Protocols}
\label{sec:ta_protocols}
When using concurrent think-aloud (TA) protocols, participants verbalize their thought processes while working on the tasks at the same time. Depending on the types of prompts or interventions administered by the study moderator, there are three common types of concurrent TA protocols: the \textit{Classic Think-Aloud protocol (CTA)}, the \textit{Speech Communication protocol (SC)}, and the \textit{Interactive Think-Aloud protocol (ITA)}. 

\textit{Classic Think-Aloud (CTA)}: CTA was established as a valid approach to studying human thinking processes by Ericsson and Simon~\cite{ericsson:1984} and later introduced into the fields of HCI and UX to study UX problems. Ericsson and Simon proposed a set of guidelines for conducting CTA: have a practice think-aloud session before the actual study session; use neutral instructions that do not direct participants to verbalize a specific type of thought process; keep prompts and interventions to a minimum by only reminding participants to ``keep talking'' if they fall into silence for a period. 

\textit{Speech Communication protocol (SC)}: Realizing the unnaturalness of thinking aloud and helping to promote thinking aloud in usability testing, Boren and Ramey found that CTA can be hard to execute by UX evaluators in practice~\cite{boren2000thinking} and proposed the SC protocol, which asked the moderator to play an active listener role by using tokens, such as “Em hmm”, “And now? ...”, in addition to ``keep talking'', to help participants think aloud. 

\textit{Interactive Think-Aloud (ITA)}: In practice, it is not uncommon that the moderator actively asks participants questions during think-aloud usability test sessions to inquire about their opinions, explanations, or suggestions~\cite{hertzum2018usability,rubin2008handbook,dumas1999practical}. This variation of the concurrent TA protocols was often referred to as interactive think-aloud (ITA). 

Since prior studies showed the pros and cons of these three TA protocols (e.g., ~\cite{olmsted2010think,olmsted2010think2,alhadreti2017intervene}), we included all the three TA protocols to understand whether these TA protocols affect native and non-native English speakers' verbalizations.

\subsection{Language Proficiency in Think-Aloud Studies}

When conducting think-aloud studies in English, prior research often recruited participants of native or fluent speakers to minimize the potential affects of language proficiency. 
We conducted a literature review and identified 16 papers that studied think-aloud protocols~\cite{andreasen2007happened,bruun2009let,cooke2010,elling2012,hertzum:2009,zhao2010keep,globalinstruction2013,remoteut2011,olmsted2010compare3,thompson2004here,haak2004ci,alhadreti2017intervene,fan2020automatic,fan2019concurrent,hertzum2015thinking,krahmer2004compare2}. Eleven of these papers did not specify the language proficiency of their participants~\cite{andreasen2007happened,bruun2009let,cooke2010,elling2012,hertzum:2009,zhao2010keep,globalinstruction2013,remoteut2011,olmsted2010compare3,thompson2004here,haak2004ci}. The five papers left mentioned that their participants were either native speakers or competent in the language that they spoke in their studies~\cite{alhadreti2017intervene,fan2020automatic,fan2019concurrent,hertzum2015thinking,krahmer2004compare2}. To the best of our knowledge, no studies have explicitly compared think-aloud verbalizations and UX problems of native and non-native speakers of a language. In this research, we sought to fill in this literature gap.

\subsection{Users' Verbalizations in Think-Aloud Studies}
\label{sec:verbalization-categories}

To understand what participants verbalize in think-aloud sessions, researchers coded participants' verbalizations and identified different \textbf{verbalization categories}.
In an early work, Bowers et al. identified five verbalization categories:
Procedure, Explanation, Reading, Design, and Others~\cite{bower1990}.
Later, Cooke conducted a study using a website and identified five similar categories: \textit{Procedure, Reading, Observation, Explanation, and Others}~\cite{cooke2010}. 
Procedure refers to participants' verbalizations that describe ``participants' current or future actions''; Reading refers to participants' verbalizations when they read any information (e.g., link labels, phrases, or sentences) from the test product;
Observation	refers to participants' verbalizations when they make remarks or observations about the test product or about themselves; Explanation refers to participants' verbalizations whey they explain their behaviors with the test product; Others	refer to verbalizations that do not fit in the above four categories. These five categories of verbalizations were later confirmed in Elling et al.'s study, in which participants used more websites than Cooke's study~\cite{elling2012}.
Although later studies divided verbalizations into more categories (e.g., ~\cite{zhao2010keep,hertzum2015thinking}), these categories could be mapped into the five-category scheme proposed in Cooke's study~\cite{cooke2010}.
Recently, Fan et al. also adopted Cooke's five-category scheme to study the correlations between verbalization categories and UX problems among native English speakers~\cite{fan2019concurrent}. 
Following these prior studies, in this study, we also adopted Cooke's categorization strategy to analyze our participants' verbalizations and investigated how these verbalization categories indicate UX problems for both non-native and native English speaking groups.  

\section{Method}
We present the details of the IRB-approved user study in the section.

\subsection{Participants}
We recruited 18 participants through social media platforms, word-of-mouth, and snowball sampling. 
All participants were undergraduate or graduate students in US universities except one who recently graduated.
One group of participants (N=9) was native English speakers from the US (8) and Canada (1), and the other group of participants (N=9) was Chinese non-native English speakers who studied in US universities.

\begin{table}[htb!]
  \caption{None-native English speakers' basic demographic information}
  \label{tab:participants}
  \begin{tabular}{ccccc}
  \toprule
 ID	&	Education level	&	English Test Score	& Proficiency Level\\
    \midrule
1		&	Graduate	&	TOEFL 91 & C1 	\\
2		&	Graduate	&	TOEFL 96 & B2 	\\
3		&	PhD	&	TOEFL 80 & B1	\\
4		&	Graduate	&	IELTS 6.5 & B2 \\
5		&	Graduate	&	TOEFL 94 & B1 	\\
6		&	Undergraduate	&	TOEFL 110 & B2 	\\
7		&	Undergraduate	&	N/A & C1 	\\
8		&	Undergraduate	&	N/A	& C2 \\
9		&	Graduate	&	TOEFL 106 & C1 	\\
    \bottomrule
\end{tabular}
\end{table}

Table~\ref{tab:participants} shows the demographic information of the non-native English speakers. In addition to the standard English test scores (e.g., TOEFL, IELTS), the researchers informally assessed participants' proficiency level based on their conversations with these participants using the Common European Frame of Reference for Languages (CEFR) guidelines~\cite{CommonEu66:online}. The definitions of the related levels were as follows: B1: intermediate; B2: upper intermediate; C1: effective operational proficiency; C2: proficiency.


\subsection{Tasks}

We chose three types of websites to increase variations of test products. One was an example of \textit{information-rich} websites\footnote{https://airandspace.si.edu/} (web 1); one was an example of \textit{e-commerce} websites\footnote{https://www.lazada.com.my/} (web 2 ); and the last one was an example of \textit{productivity-enhancement} websites\footnote{https://basecamp.com/} (web 3). 

\begin{figure}[htb!]
  \centering
  \includegraphics[width=0.6\linewidth]{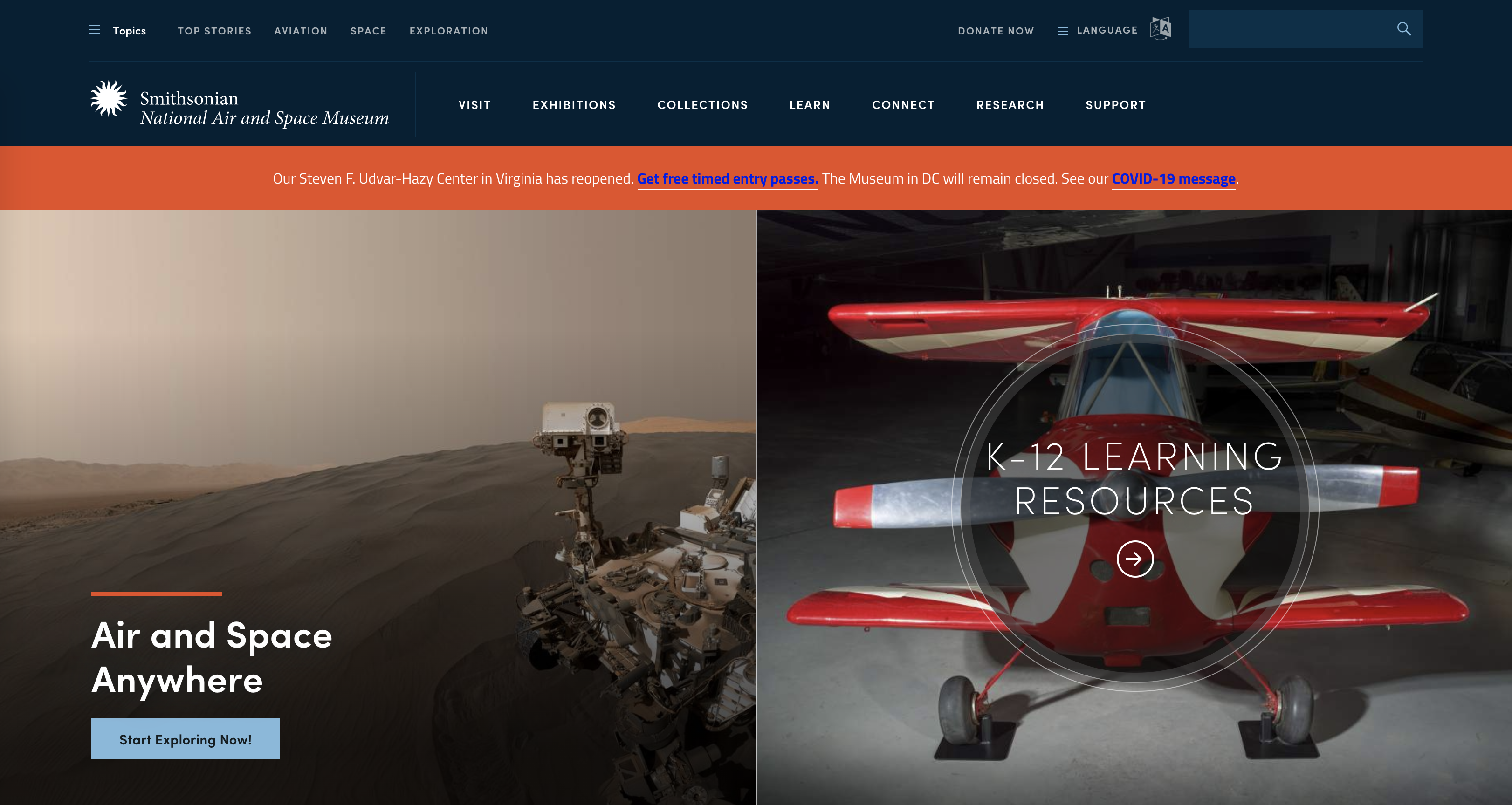}
  \caption{Web 1: an Air and Space Museum (an information-rich website)}
  \Description{Web 1: an Air and Space Museum}
  \label{fig:museum}
\end{figure}

\begin{figure}[htb!]
  \centering
  \includegraphics[width=0.6\linewidth]{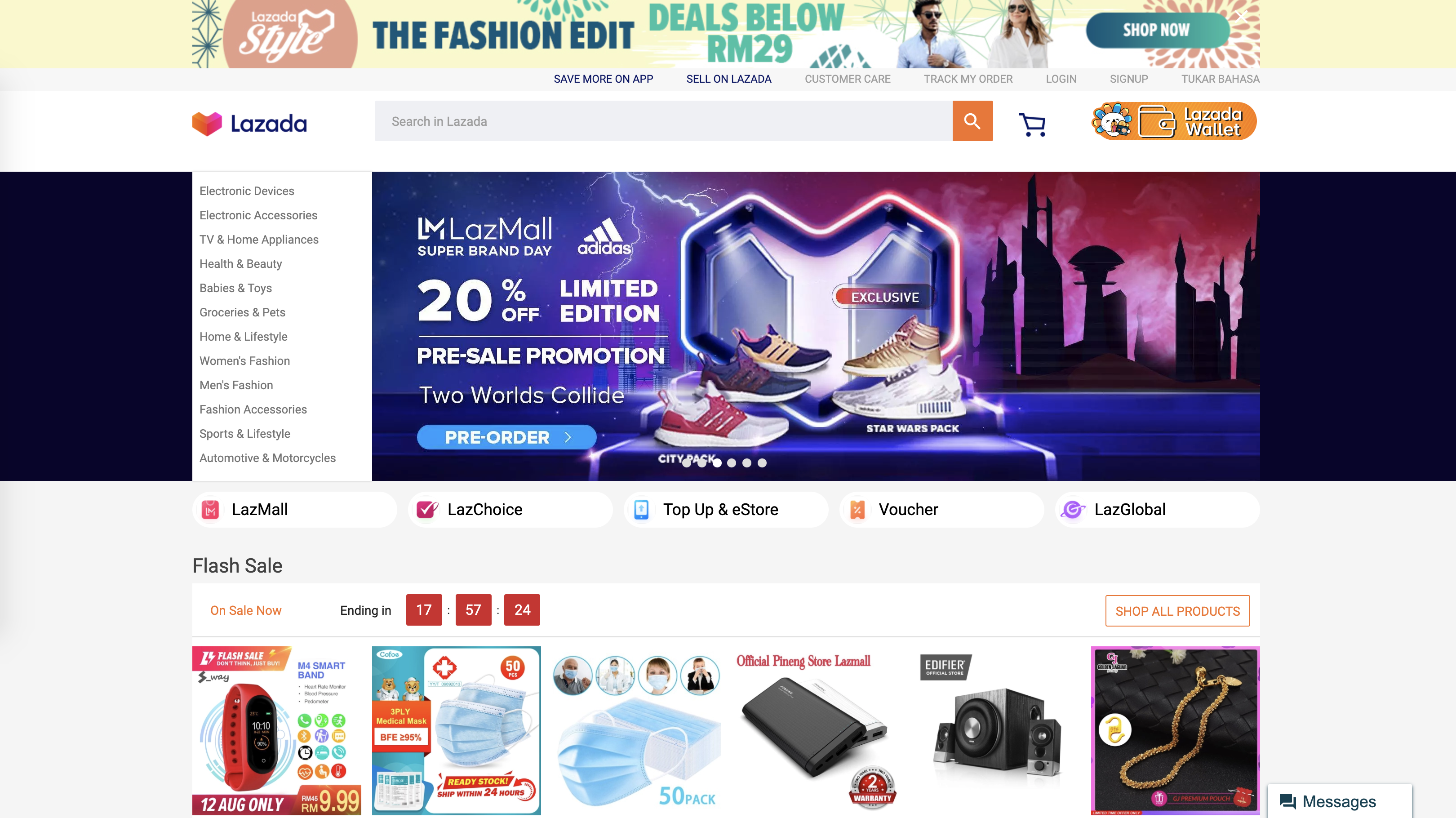}
  \caption{Web 2: Lazada (an E-commerce website)}
  \Description{Web 2: The homepage of Lazada (an E-commerce website)}
  \label{fig:lazada}
\end{figure}

\begin{figure}[htb!]
  \centering
  \includegraphics[width=0.6\linewidth]{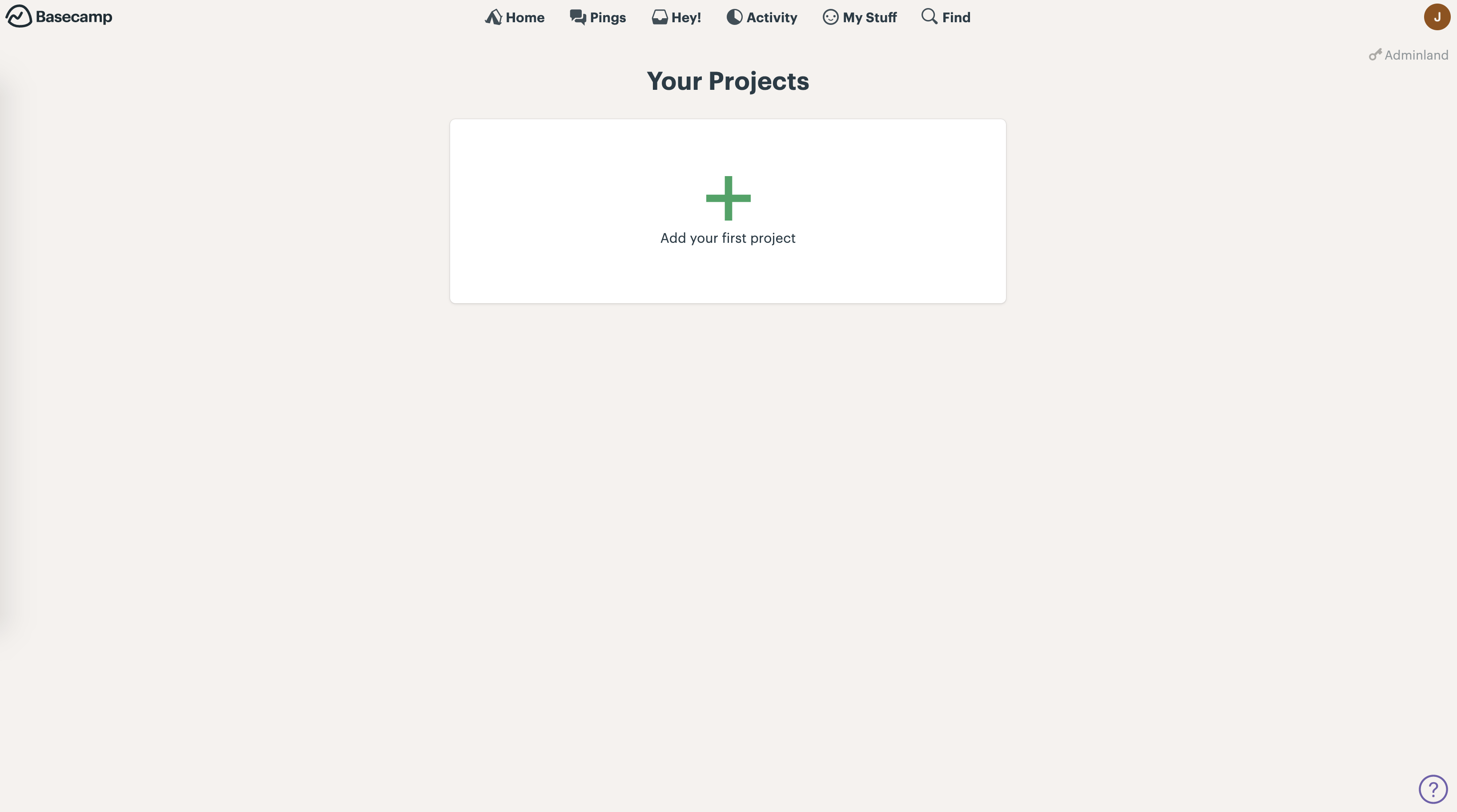}
  \caption{Web 3: Basecamp (an online teamwork tool)}
  \Description{Web 3: Basecamp (an online teamwork tool)}
  \label{fig:basecamp}
\end{figure}
Figure~\ref{fig:museum} shows the homepage of the museum website (Air and Space Museum) website.
Figure~\ref{fig:lazada} shows the homepage of the e-commerce website (Lazada).
Figure~\ref{fig:basecamp} shows the homepage of the teamwork tool website (Basecamp).
We chose these websites because 1) they represent common types of websites one would use in their daily life; 2) we conducted heuristic evaluation using both Nielsen heuristics~\cite{nielsen:1994} and Norman principles~\cite{norman2013design} and found that all of these websites contained UX problems; and 3) none of the participants used these websites prior to the study. 

We designed corresponding tasks that would require participants to use the features of the websites that contained UX problems.
Table~\ref{tab:tasks} shows the websites and the corresponding tasks used in the study.

\begin{table*} [htb!]
  \caption{Tasks for participants to complete on each website}
  \label{tab:tasks}
  \begin{tabular}{p{3.3cm}|p{13.7cm}}
    \toprule
  Websites &	Tasks	\\
    \hline
Web 1 (Museum, an information-rich website)	&	1) Find out how much it will cost to enter the museum, and days of closure; 
2) Find out what things you can do in the museum; 3) Find out what exhibitions are on view currently; 4) Find out where the exhibition (Boeing Milestones of Flight Hall) is on the museum’s map, and find out what you will see in this exhibition; 5) Find out if there is an audio guide in the museum.\\
	\hline
Web 2 (Lazada, an e-commerce website)	&	1) Find out a smartwatch that meets all the 3 requirements: can be used underwater; can track your heart rate; has great customer ratings; 2) Find out what other customers said good about the watch you just found; 3) Find out where you can ask the seller questions; 4) Find out other products from the same seller; 5) Find out where you can know the shipping status of your order;	\\
	\hline
Web 3 (Basecamp, an online teamwork tool)	&	1) Create a project named "birthday party"; 2) Invite two persons into the project; 3) Share the idea of decoration you like (a link) with them; 4) Assign tasks to each person; 5) Delete the project.\\
    \midrule
\end{tabular}
\end{table*}

\subsection{Procedure}

We conducted the user study with participants online using Zoom. We screen- and audio-recorded study sessions. The study lasted for about an hour.

Upon signing the consent form, the moderator explained to participants that they would complete tasks on three websites and think aloud during the process. To help participants better understand how to think aloud, the moderator first explained verbally how to think aloud and then showed a think-aloud demo video provided by the Nielsen Norman group~\cite{NNGroup}, which shows a user working on a website while thinking aloud at the same time. Afterward, the moderator asked participants to practice thinking aloud on a website, which was not one of the test websites, to complete a task. Next, the moderator asked participants to work on the three test websites to complete corresponding tasks in Table~\ref{tab:tasks} while thinking aloud at the same time. For each website, participants used one of the three TA protocols (i.e., CTA, SC, ITA) as explained in Sec~\ref{sec:ta_protocols}. The order of TA protocols and the test websites were counterbalanced using the Latin square design in Table~\ref{tab:latin}.  

\begin{table*}[htb!]
  \caption{The study design with three TA protocols (CTA, SC, ITA) and test websites counter-balanced for two language groups.}
  \label{tab:latin}
  \begin{tabular}{p{3.2cm}|p{3.2cm}|p{1.6cm}|p{1.6cm}|p{1.6cm}}
    \toprule
   
 Non-Native speakers' IDs	&	Native spearkers' IDs	&  \multicolumn{3}{c}{TA Protocols and websites} \\
    \hline
P1	&	P10	&	CTA (web 1) &	SC (web 2)	&	ITA (web 3)	\\ \hline
P2	&	P11	&	SC (web 1) &	ITA	(web 2) &	CTA	(web 3)\\ \hline
P3	&	P12	&	ITA (web 1) &	CTA	(web 2) &	SC (web 3)	\\ \hline
P4	&	P13	&	CTA (web 2) &	SC (web 3)	&	ITA (web 1)	\\ \hline
P5	&	P14	&	SC (web 2)	&	ITA(web 3)	&	CTA (web 1)	\\ \hline
P6	&	P15	&	ITA	(web 2) &	CTA (web 3)	&	SC (web 1)	\\ \hline
P7	&	P16	&	CTA (web 3) &	SC (web 1)	&	ITA (web 2)	\\ \hline
P8	&	P17	&	SC (web 3)	&	ITA(web 1)	&	CTA (web 2)	\\ \hline
P9	&	P18	&	ITA	(web 3) &	CTA (web 1)	&	SC (web 2)	\\
    \bottomrule
\end{tabular}
\end{table*}

When conducting CTA sessions, the moderator followed Ericcson and Simon's guidelines and did not use any prompts and only reminded participants to ``keep talking'' if they fell into silence for more than 10 seconds. 
When conducting SC sessions, the moderator followed the guidelines put forward by Boren and Ramey~\cite{boren2000thinking} and played an active listener's role by saying phrases such as ``Em, hmm'', ``uh-huh'', ``and now?'', and ``keep talking'' to encourage participants to think aloud.
When conducting ITA sessions, the moderator actively probed participants using five types of prompts that were derived from prior studies~\cite{alhadreti2017intervene,hertzum2018usability,zhao2010keep}: \textit{Clarifying intentions} (“What are you looking for?”), \textit{Seeking explanations} (“Could you tell me why you did that?”), \textit{Seeking opinions} (“What do you think of it?”), \textit{Seeking suggestions} (“What redesign do you suggest?”), and \textit{Seeking user expectation} (“what do you expect to be there?”). 

After participants completed each task, they were asked to fill in the NASA Task Load Index (TLX) form to measure their perceived task load of completing the task while thinking aloud.  



\section{Analyses}


\subsection{Categorizing Think-aloud (TA) Verbalizations}
\label{sec:categorizing_verbalizations}


We first used an automatic transcribing tool, Otter.ai~\cite{OtterVoi54:online}, to transcribe session recordings and then manually checked the transcriptions to correct errors. After this process, we had all participants' think-aloud verbalizations.

We followed a similar process used in the literature~\cite{fan2019concurrent,cooke2010,elling2012,zhao2010keep} to review each think-aloud session recording and segment it into smaller segments based on pauses in the participant's verbalizations and the semantics of the verbalizations. A segment could include sentences, phrases, or single words.  

Next, for each segment, two UX researchers independently reviewed the corresponding verbalizations and assigned it a \textit{verbalization category} label using Cooke's five-category scheme~\cite{cooke2010} as explained in Sec~\ref{sec:verbalization-categories}. This scheme was widely adopted by prior studies~\cite{elling2012,fan2019concurrent,hertzum2015thinking,Fan2021OlderAdults}. The five verbalization categories were: \textit{Procedure}, \textit{Reading}, \textit{Observation}, \textit{Explanation}, and \textit{Others}. Table~\ref{tab:category} shows the categories, their definitions, and examples from our participants' think-aloud verbalizations.

\begin{table*}[htb!]
  \caption{Verbalization categories, definitions and examples from the study.}
  \label{tab:category}
   \begin{tabular}{p{1.6cm}|p{7cm}| p{7.5cm}}
    \toprule
Categories	&	Definitions & Examples	\\
    \midrule
Procedure	&	Describe their current or future actions & \textit{"I'll start with sports and lifestyle."} \textit{"Where should I go?"}\\  
\hline
Reading	&	Read any information (e.g., link labels, phrases, or sentences) from the test product	& \textit{"The all-in-one toolkit for working remotely."}\\
\hline
Observation	&	Make an observation or a remark about the test product or themselves (e.g., comments, feelings) &	\textit{"Looks like it's selling everything under the sun."} \textit{"So I think that makes it a little bit more overwhelming."}\\
\hline
Explanation	&	Explain their behaviors on the test product &	\textit{"I think underwater was probably a bad term, maybe I had to search for waterproof."}\\
\hline
Others	&	Verbalizations that do not fit in the above four categories: task-related (e.g., read task descriptions or ask questions about tasks); verbal fillers (e.g, Um, Ah, alright) & \textit{"I want to buy a smartwatch to track my swimming exercise [task]." (note that this participant was paragraphing the task)} \textit{"Alright."} \textit{"Let's see."}\\
    \bottomrule
\end{tabular}
\end{table*}

After each researcher finished assigning category labels for each segment independently, they reviewed their category labels together to gain a consensus on the labels. If there was a disagreement, they explained their rationales for their labels, discussed with each other,  and consolidated the category labels.





\subsection{Identifying UX Problems}
For each verbalization segment, two UX researchers followed the same procedure of assigning verbalization category labels in Sec~\ref{sec:categorizing_verbalizations} to determine whether the user encountered a problem and to assign a binary problem label (0: no problem; 1: problem). 
Each segment with a problem label ``1'' represents ``a moment in which a user encountered a problem'' and was referred to as a \textbf{``problem encounter.''}
Because different ``problem encounters'' might be caused by the same underlying UX problem, two UX researchers further reviewed these ``problem encounters'' and combined the problem encounters caused by the same underlying problem, which was referred to as an \textbf{``actual problem''}. 
As a result, the number of ``actual problems'' would be less or equal than the number of ``problem encounters.'' 

Furthermore, two researchers followed the same procedure to assess the severity of each ``actual problem'' using Nielsen's guidelines~\cite{Severity64:online}. The definitions of the five severity levels are as follows: level 0 means ``no usability problem''; level 1 means ``cosmetic problem''; level 2 means ``minor usability problem that should be given low priority''; level 3 means ``major usability problem that should be given high priority''; level 4 means ``usability catastrophe that imperative to fix before product can be released''~\cite{Severity64:online}.

\section{Results}
\label{sec:results}
\subsection{Verbalization Categories}

\begin{table}[ht]
  \caption{Number (percentage) of verbalization segments in each verbalization category for each language group.}
  \label{tab:numGroup}
  \begin{tabular}{ccc}
    \toprule
  Category	&	Native	&	Non-native\\
    \midrule
Observation 	&	752 (34.2\%)	&	758 (33.5\%)	\\
Procedure	&	619 (28.1\%)	&	793 (35.0\%)	\\
Others	&	434 (19.7\%)	&	336 (14.8\%)	\\
Reading	&	286 (13.0\%)	&	207 (9.1\%)	\\
Explanation	&	111 (5.0\%)	&	172 (7.6\%)	\\
\hline
Total	&	2202 (100.0\%)	&	2266 (100.0\%)	\\
    \bottomrule
\end{tabular}
\end{table}
\subsubsection{Verbalization Categories grouped by \textbf{Language Groups}}
Table ~\ref{tab:numGroup} shows the number and percentage of segments in each verbalization category for native and non-native English speaking participants. 
Results suggest that the general trends for the two language groups were similar. 
Specifically, two most frequently verbalized categories (i.e., Procedure and Observation) and three least frequently verbalized categories (i.e., Others, Reading, and Explanation) were the same for both native and non-native English speaking participants. 

One difference was that, for native English speakers, Observation was the most frequently verbalized category followed by Procedure.
For non-native English speakers, Procedure was the most frequently verbalized category, followed by Observation. In other words, compared to native speakers, non-native speakers verbalized a relatively higher percentage of what they were doing (i.e., Procedure) than what they were remarking (i.e., Observation)

\subsubsection{Verbalization Categories grouped by \textbf{TA Protocols}}
\begin{table}[tbh]
  \caption{Number (percentage) of verbalization segments in each verbalization category for each TA protocol.}
  \label{tab:numratioProtocol}
  \begin{tabular}{cccc}
    \toprule
 Category	&	CTA	&	SC	&	ITA	\\
    \midrule
Procedure	&	508 (34.6\%)	&	416 (33.2\%)	&	488 (27.9\%)	\\
Observation 	&	460 (31.4\%)	&	395 (31.5\%)	&	655 (37.5\%)	\\
Others	&	267 (18.2\%)	&	242 (19.3\%)	&	261 (14.9\%)	\\
Reading	&	168 (11.5\%)	&	126 (10.0\%)	&	199 (11.4\%)	\\
Explanation	&	64 (4.4\%)	&	75 (6.0\%)	&	144 (8.2\%)	\\
\hline
Total	&	1467 (100.0\%)	&	1254 (100.0\%)	&	1747 (100.0\%)	\\
    \bottomrule
\end{tabular}
\end{table} 

Table ~\ref{tab:numratioProtocol} shows the number and ratio of verbalization segments in each category by TA protocols. 
For all TA protocols, Observation and Procedure were the most frequently appeared two categories, followed by Others, Reading, and Explanation. 

While CTA and SC had slightly higher percentages of Procedure than Observation, ITA had a slightly higher percentage of Observation than Procedure.
In other words, with the CTA or SC protocol, participants tended to verbalize what they were doing (i.e., Procedure) more often than to make remarks (i.e., Observation). In contrast, under the ITA protocol, participants tended to make remarks (i.e., Observation) more often than verbalizing what they were doing (i.e. Procedure).

\subsubsection{Effects of Language Groups and TA Protocols}
\label{sec:effects_language_groups_TA_protocols}
To further understand the effects of language groups and TA protocols, we performed three-way ANOVA with both \textit{TA protocols} and \textit{verbalization categories} as within-subjects factors and \textit{language groups} as the between-subjects factor. Results show 1) no significant difference for the \textit{language groups} ($F(1, 16) = 0.062, p=0.806, \eta_{p}^{2}=0.004$); 2) significant differences for \textit{TA protocols} ($F(2, 32) = 4.621, p=0.017, \eta_{p}^{2}=0.223$) and \textit{verbalization categories} ($F(4, 64) = 44.440, p < 0.0001, \eta_{p}^{2}=0.71$). We further performed Sheffe Post-hoc analysis for \textit{verbalization categories} and \textit{TA protocols}.

For the \textit{native language group}, Scheffe Post-hoc analysis found significant differences for the following pairs: (Explanation, Observation), (Explanation, Procedure), (Explanation, Others), (Reading, Observation), and (Reading, Procedure). 
Similarly, For the \textit{non-native language group}, Scheffe Post-hoc analysis found significant differences for the following pairs: (Explanation, Observation), (Explanation, Procedure), (Others, Observation), (Others, Procedure), (Reading, Observation), and (Reading, Procedure). In other words, the significant differences were between the least and most frequently appeared categories for both native and non-native language groups. 

For \textit{TA protocols}, Scheffe Post-hoc analysis showed that they did not have any significant effect on all verbalization categories except the Explanation category. Specifically, ITA had significantly more Explanation than CTA and SC. In other words, participants tended to explain their behaviors more often in ITA than in CTA or SC.

\subsection{UX Problems}

\subsubsection{UX Problems and Examples}

The test websites were used as vehicles to answer our RQs, which focused on understanding non-native and native English speakers' think-aloud verbalizations and the correlations between the verbalizations and UX problems. Nonetheless, we present example UX problems, the usability heuristics violated~\cite{nielsen:1994}, and participants' think-aloud verbalizations in Table~\ref{tab:UXproblems_examples} to better contextualize the results presented in the rest of Sec.~\ref{sec:results}.

\begin{table*}[htb!]
  \caption{The UX problems with usability heuristics violated and the example think-aloud verbalizations.}
  \label{tab:UXproblems_examples}
  \begin{tabular}{p{6cm}|p{11cm}}
    \toprule
 UX Problems (\textit{Usability heuristics violated}~\cite{nielsen:1994}) & Problem description with think-aloud verbalizations	\\
    \midrule
The design did not speak users' language or failed to match users' mental model.  \textit{(Match between system and the real world)}	&	

\textbf{Web 2 (Lazada)}: The options were not organized in a natural and logical order. It took P2 a while to find the target function, and she verbalized, \textit{"Where's my function? I feel like there are too many (options) here."}
	\\
\hline


    The design failed to provide users with effective error messages that could indicate problems and suggest solutions   (\textit{Help users recognize, diagnose, and recover from errors}) 	&	\textbf{Web 1 (Museum)}: The navigation disappeared when P12 tried to move the mouse to the sub-navigation. He verbalized, \textit{"Here, visit, Oops...[navigation bar disappears], visit."}	\\
\hline
The design failed to keep users informed about what is going on through appropriate feedback. \textit{(Visibility of system status)}	&	\textbf{Web 1 (Museum)}: P18 was confused about the map and verbalized, \textit{"I don't know where exactly in the map is it."}

\textbf{Web 2 (Lazada)}: The workspace icon was not self-explainable and needed additional explanation. P13 verbalized, \textit{"Okay, so pretty blank here. Doesn't really tell you what you can do."}	\\
\hline
The design failed to prevent problems from happening in the first place. \textit{(Error prevention)}  &	\textbf{Web 3 (Basecamp)}: The accent color mislead users to make mistakes. Instead of sending the project out as required, P4 accidentally saved the project as a draft because the "Draft" button was green and the "Post" button was white. He verbalized, \textit{"Send as a draft... Oh, No. Post this."}	\\
    \bottomrule
\end{tabular}
\end{table*}

\subsubsection{UX Problems grouped by \textbf{Language Groups}}

Table ~\ref{tab:problem-encounters} shows the number of problem encounters and actual problems for two language groups respectively.
To reiterate, the number of actual problems was 15 and 18 for native and non-native English participants respectively. 
\begin{table}[ht]
  \caption{Number of problem encounters and actual problems for each language group.}
  \label{tab:problem-encounters}
  \begin{tabular}{ccc}
    \toprule
 Group	&	Problem encounters &	Actual problems\\
    \midrule
Native	&	35	 &  15\\
Non-native	&	41	 & 18\\
\hline
Total	&	76	 & 19 \\
    \bottomrule
\end{tabular}
\end{table}

We further analyzed \textit{common} actual problems in both language groups and the \textit{unique} actual problems to each language group. Among the 19 actual problems, 14 were identified as common problems for both language groups and five were unique problems to each group. Four out of the five unique problems were of the lowest severity level 1 (i.e., cosmetic problem~\cite{Severity64:online}) and the rest one was also of low severity level 2 (i.e., minor problem~\cite{Severity64:online}). In other words, both native and non-native speakers' think-aloud verbalizations were equally effective in identifying UX problems of high severity levels with only minor differences in revealing UX problems of low severity levels. 

\subsubsection{UX Problems grouped by \textbf{TA Protocols}}
We also counted the number of problems encounters in each TA protocol. Table ~\ref{tab:problems-by-protocols} shows the number of problem encounters and actual problems identified for each TA protocol. 

\begin{table}[ht]
  \caption{Number of problem encounters and actual problems for each TA protocol.}
  \label{tab:problems-by-protocols}
  \begin{tabular}{ccc}
    \toprule
  Group	&	Problem encounters &	Actual problems\\
    \midrule
CTA	&	25	& 12\\
SC	&	15	& 12\\
ITA	&	36	& 15\\
\hline
Total	&	76	 & 19 \\
    \bottomrule
\end{tabular}
\end{table}

\subsubsection{Effects of Language Groups and TA Protocols}

\textit{For the number of problem encounters}, ANOVA analysis found no significant effect of the \textit{language groups} ($F(1,16)=0.252, p=0.618, \eta_{p}^{2}=0.02$). This suggests native and non-native English speakers' verbalizations do not have significant difference in identifying UX problems.
While ANOVA analysis found a significant effect of \textit{TA protocols} ($F(2,32)=5.122, p=0.034, \eta_{p}^{2}=0.24$), Scheffe post-hoc analysis did not find significant difference among any pairs of TA protocols. This suggests that the three types of TA protocols do not have significant difference in identifying UX problems.

Similarly, for \textit{the number of actual problems}, ANOVA results also found no significant effect of \textit{language groups} ($F(1,16)=0.313, p=0.583, \eta_{p}^{2}=0.019$) or \textit{TA protocols} ($F(2,32)=2.582, p=0.091, \eta_{p}^{2}=0.138$). This again suggests there are no significant differences in identifying UX problems between language groups or between TA protocols.



\subsection{Correlations between Verbalization Categories and UX Problems}
\label{sec:correlations_verbalizations_problems}
To understand how each verbalization category is indicative of UX problems, we counted the number of segments in each \textit{verbalization category} (i.e., Procedure, Reading, Observation, Explanation, and Others) that were associated with a UX problem. We then grouped them by \textit{language groups} and \textit{TA protocols}. Table~\ref{tab:verbAndProb} and Table~\ref{tab:verbAndProbByProtocol} showed the results of the number of segments in each verbalization category indicating UX problems for language groups and for three TA protocols  respectively. 

\begin{table}[htb!]
  \caption{Number (percentage) of segments in each category indicating UX problems for each language group.}
  \label{tab:verbAndProb}
  \begin{tabular}{ccc}
    \toprule Category	&	Native	&	Non-native\\
    \midrule
Observation 	&	57 (58.8\%)	&	66 (65.3\%)	\\
Procedure	&	14 (14.4\%)	&	22 (21.8\%)	\\
Explanation	&	14 (14.4\%)	&	10 (9.9\%)	\\
Others	&	12 (12.4\%)	&	3 (3.0\%)	\\
Reading	&	0 (0.0\%)	&	0 (0.0\%)	\\
\hline
Total	&	97 (100.0\%)	&	101 (100.0\%)	\\
    \bottomrule
\end{tabular}
\end{table}
\begin{table}[htb!]
  \caption{Number (percentage) of segments in each category indicating UX problems for each TA protocol.}
  \label{tab:verbAndProbByProtocol}
  \begin{tabular}{cccc}
    \toprule
  Category	&	CTA	&	SC	&	ITA	\\
    \midrule
Observation 	&	31 (55.4\%)	&	31 (77.5\%)	&	61 (59.8\%)	\\
Procedure	&	13 (23.2\%)	&	5 (12.5\%)	&	18 (17.6\%)	\\
Others	&	9 (16.1\%)	&	3 (7.5\%)	&	3 (2.9\%)	\\
Explanation	&	3 (5.4\%)	&	1 (2.5\%)	&	20 (19.6\%)	\\
Reading	&	0 (0.0\%)	&	0 (0.0\%)	&	0 (0.0\%)	\\
\hline
Total	&	56 (100.0\%)	&	40 (100.0\%) & 	102 (100.0\%)\\
    \bottomrule
\end{tabular}
\end{table} 
\subsubsection{Correlations organized by \textbf{Language Groups}}
As shown in Table~\ref{tab:verbAndProb}, the number of segments of each verbalization category related to a UX problem was similar for two language groups: 97 verbalization segments related to a problem in the native language group and 101 in the non-native language group.
Moreover, the order of the verbalization categories indicating UX problems (from the most to the least) was also the same for both native and non-native language groups. These results suggest that language groups do not significantly affect how verbalization categories indicate UX problems. 


\subsubsection{Correlations organized by \textbf{TA Protocols}}

As shown in Table ~\ref{tab:verbAndProbByProtocol}, the order of the verbalization categories from the most to the least indicative of UX problems was the same for CTA and SC. While ITA followed a similar trend as CTA and SC, Explanation in ITA was more indicative of problems than in CTA or SC. Moreover, while CTA and SC had a similar number of segments related to a problem, ITA had more segments related to problems than CTA or SC.   




\subsection{Task Load}
\subsubsection{Task Load by \textit{language groups}}
\begin{table}[htb!]
  \caption{Average ratings of the six scales of the NASA Task Load Index (TLX) grouped by language groups.}
  \label{tab:taskload_group}
  \begin{tabular}{cccc}
    \toprule
TLX scales (range: 1---21)	&	Native	&	Non-native		\\
    \midrule
Mental demand	&	4.2	&	5.4		\\
Physical demand	&	3.3	&	4.0		\\
Temporal demand	&	4.3	&	5.6	\\
Performance	&	15.1	&	16.0	\\
Effort	&	4.8	&	5.6		\\
Frustration	&	4.1	&	3.2		\\
    \bottomrule
\end{tabular}
\end{table}
Table~\ref{tab:taskload_group} shows the average ratings of the six scales of the NASA Task Load Index (TLX) grouped by language groups. The higher the score, the higher the task load in that scale, except for the performance scale. Results show that: 1) overall, both native and non-native English speaking participants felt that thinking aloud was not much demanding and their performance was high; 2) Compared to native English speaking participants, non-native English speaking participants felt that thinking aloud was relatively more mentally, physically, and temporally demanding, more effortful, and more frustrated. However, the differences in each scale between the two language groups were not significant based on ANOVA.


\begin{table}[htb!]
  \caption{Average ratings of the six scales of the NASA Task Load Index (TLX) grouped by three TA protocols.}
  \label{tab:taskload_protocol}
  \begin{tabular}{ccccc}
    \toprule
TLX scales (range: 1---21)	&	CTA	&	SC	&	ITA	\\
    \midrule
Mental demand	&	4.8	&	4.9	&	4.8	\\
Physical demand	&	3.6	&	3.4	&	3.9		\\
Temporal demand	&	4.9	&	4.7	&	5.4		\\
Performance	&	15.3	&	16.4	&	14.9	\\
Effort	&	4.7	&	5.1	&	5.8		\\
Frustration	&	3.5	&	3.1	&	4.3		\\
    \bottomrule
\end{tabular}
\end{table}
\subsubsection{Task Load by \textit{TA Protocols}}
Table~\ref{tab:taskload_protocol} shows the average ratings of the six scales of the NASA Task Load Index (TLX) grouped by three TA protocols. Results show that: 1) regardless of TA protocols, participants felt that thinking aloud was not much demanding and their performance was high; 2) the differences in all scales between three protocols were not significant based on ANOVA.

\section{Discussion}
\subsection{The Effect of Language Groups}
We discuss the effect of language groups on three aspects: \textit{the proportions of verbalization categories}, \textit{UX problems}, and \textit{the correlations between verbalization categories and UX problems}.

\textbf{Proportions of Verbalization Categories}: Our analysis showed no significant difference in the proportion of each verbalization category for two language groups. Moreover, the proportions of five categories also followed a similar trend for two language groups. Observation and Procedure were the most frequently appeared two categories, followed by Explanation, Others, and Reading. 

While there was no statistical significance, the non-native language group did seem to verbalize a relatively higher proportion of Procedure (i.e., what they were doing) and a relatively lower proportion of Reading (i.e., what they saw) than the native language group. 

Because both Procedure and Reading were all level-1 and level-2 types of verbalizations according to Ericsson and Simon's definition~\cite{ericsson:1984}, we combined these two categories and found that the proportion of the two categories together was similar for the two language groups: 41.1\% and 44.1\% for native and non-native language groups.
To better compare our results with prior work, we further combined the verbalizations of each category across all our participants. Table~\ref{tab:comparedwithotherstudies} shows the result.
It showed that our study found similar proportions of Procedure and Reading, Observation, and Explanation as Elling et al.~\cite{elling2012}, Zhao et al.~\cite{zhao2014impact}, and Fan et al.~\cite{fan2019concurrent}, but there was a difference between ours and Cooke's study~\cite{cooke2010} or Fan et al.'s study~\cite{Fan2021OlderAdults}.
The difference might be due to differences in the test products, TA protocols used, and the participants. While Cooke only used CTA in their study~\cite{cooke2010}, we used three types of TA protocols. While we tested with websites only, Fan et al.~\cite{Fan2021OlderAdults} tested with both physical and digital products. Furthermore, while Fan et al.~\cite{Fan2021OlderAdults} focused on older adults, our participants were all young adults. Future work should conduct more-controlled studies with the same set of products, participants, and study procedures to better understand how test products and participants' age and other backgrounds might affect think-aloud verbalizations.  

\begin{table} [htb!]
  \caption{Proportions of verbalization segments in each study. N/A means the category was not used in the study. Zhao et al.'s research~\cite{zhao2014impact} used two TA protocols similar to CTA and ITA in this study.}
  \label{tab:comparedwithotherstudies}
  \resizebox{0.9\linewidth}{!}{
  \begin{tabular}{ccccc}
    \toprule
 Studies	&	Procedure	&	Observation	&	Explanation	&	Others	\\ 
 & Reading & & & \\
    \midrule
Cooke~\cite{cooke2010}	&	77\%	&	10\%	&	5\%	&	8\%	\\
Elling et al.~\cite{elling2012}	&	40\%	&	34\%	&	7\%	&	19\%	\\
Zhao et al. CTA~\cite{zhao2014impact}	&	70.3\%	&	20.1\%	&	9.6\%	&	N/A	\\
Zhao et al. ITA~\cite{zhao2014impact}	&	49.9\%	&	33.8\%	&	16.3\%	&	N/A	\\
Fan et al.~\cite{fan2019concurrent}	&	56.3\%	&	37.6\%	&	5.9\%	&	N/A	\\
Fan et al.~\cite{Fan2021OlderAdults}	&	31.2\%	&	62.5\%	&	3.4\%	&	2.9\%	\\
Our current study	&	42.6\%	&	33.6\%	&	6.3\%	&	17.2\%	\\
    \bottomrule
\end{tabular}
}
\end{table}










\textbf{UX Problems}: 
Our analysis results show that the language groups did not significantly affect either the number of problem encounters or the number of actual problems. The only difference was in the identification of some low severity UX problems. The implication is that both native and non-native English participants are equally effective in helping locate common and severe UX problems in think-aloud usability testing.

\textbf{Correlations}: 
Results show that correlations between verbalization categories and UX problems followed similar trends for both native and non-native language groups. The categories ranged from the most to the least indicative of UX problems were: Observation, Procedure, Explanation, Others, and Reading.

\subsection{The Effect of TA Protocols}
Similarly, we discuss the effect of three TA protocols on the same three aspects: the proportions of verbalization categories, UX problems, and the correlations between verbalization categories and UX problems.

\textbf{Proportions of Verbalization Categories}:
TA protocols did not have a significant effect on all categories except Explanation. ITA had significantly more proportions of Explanation than CTA and SC.
This difference was likely because the moderator prompted the participants to verbalize more by asking for explanations, opinions, and suggestions when using the ITA protocol. In contrast, such prompting behavior was forbidden in both CTA and SC protocols.  

\textbf{UX Problems}: TA protocols also did not have a significant effect on the number of actual problems. While TA protocols were shown to have a significant effect on the number of problem encounters, post-hoc analysis did not find a significant difference among TA protocols. This suggests that all TA protocols were equally effective in identifying UX problems.

Although there was no statistical significance, ITA found a higher number of problem encounters than CTA and SC, according to Table~\ref{tab:problems-by-protocols}.
To better understand this difference, we examined the severity of the problems found by each TA protocol. 
Four out of the five problems that were only found by ITS were all of the lowest severity level (i.e., cosmetic problems~\cite{Severity64:online}). Cosmetic problems were mostly related to none-essential and nice-to-have features, such as changing the cursor icon when hovering over a clickable element. In other words, all protocols were equally effective in identifying severe UX problems.  
Furthermore, this finding was also consistent with previous research, which also found CTA, SC, and ITA found a similar number and type of problems~\cite{alhadreti2017intervene}. 


\textbf{Correlations}:
Correlations between Verbalization Categories and UX Problems followed similar trends for three types of TA protocols.
The Observation category was the most indicative of problems of all categories. This was consistent with prior findings on CTA~\cite{fan2019concurrent}. 
One difference among the TA protocols was that Explanation in ITA seemed to be more indicative of problems than in CTA or SC.
This difference was likely because ITA had a higher proportion of verbalizations in the Explanation category than CTA or SC.

\subsection{Design Implications}
In this research, we took a first step to uncover similarities and differences in think-aloud verbalizations, the UX problems, and their correlations between \textit{two English language groups} (i.e., native and non-native English speakers) in \textit{three think-aloud protocols} (i.e., CTA, SC, ITA). Based on the findings, we discuss three design implications (DIs).

\textbf{DI1: None-native English speakers can be as effective as native English speakers to help identify UX problems in Think-Aloud Usability Testing}. Our findings show that the verbalization patterns (i.e., verbalization categories and how they indicate UX problems) observed in native English speakers in prior research~\cite{cooke2010,elling2012,zhao2010keep,fan2019concurrent,Fan2021OlderAdults} were mostly applicable to non-native English speakers whose English proficiency was on or above the intermediate level~\cite{CommonEu66:online}.
Thus, UX practitioners could consider to enroll English speakers to their think-aloud usability testing for identifying UX problems without needing to worry about whether they are native or non-native speakers as long as their English reaches an intermediate level~\cite{CommonEu66:online}. 

\textbf{DI2: Three concurrent think-aloud protocols (i.e., CTA, SC, and ITA) are equally effective in identifying severe UX problems}. Our results suggest that the three types of think-aloud protocols (i.e., CTA, SC, and ITA) do not significantly affect the number of UX problems. While ITA might be able to identify slightly more problems, these extra problems are often of low severity levels. In contrast, ITA requires significantly more effort from the study moderator. The moderator has to constantly probe participants with different types of questions, which increases her work load. As a result, we suggest UX practitioners stick with Ericcson and Simon's classic think-aloud protocol (CTA)~\cite{ericsson:1984}, which not only minimizes the effort from the moderator, who only needs to remind participants to ``keep talking'' if they fall into silence, but also is able to uncover a similar set of important UX problems as the other two TA protocols (i.e., SC, ITA).

\textbf{DI3: Non-native English speakers' verbalizations could be used to increase the amount of training data to build artificial intelligence (AI) models to detect UX problems automatically}. 
Our results show that similar verbalization patterns indicating UX problems were found in both non-native English speakers and native English speakers. 
For example, as indicated in Sec~\ref{sec:correlations_verbalizations_problems}, for both language groups, the most verbalized categories were Observation and Procedure, which were also more indicative of UX problems than other categories. Such patterns could be used to build AIs to help UX practitioners better identify UX problems.
For example, Fan et al. showed that verbalization category label was an effective feature to train AIs to detect UX problems automatically along with other textual and acoustic features, such as sentiment, speech rate, pitch and loudness~\cite{fan2020automatic}. Meanwhile, verbalization category information could also be visualized to direct UX evaluators' attention to segments of a TA usability test video that are more likely associated with a UX problem~\cite{fan2020vista}.

One key challenge in building such AIs or AI-assisted tools is to gather a large amount of training data. This is even more challenging if UX practitioners have to focus on recruiting native English speakers due to the concerns of the effect of participants' language proficiency on identifying UX problems. Indeed, outside of a few countries where English is the first language, it is almost impossible to recruit a sufficiently large number of native English speakers. Luckily, many countries around the world have a critical mass of regular non-native English speakers. Our research provides evidence that non-native English speakers' TA verbalizations can be as effective as those of native speakers for identifying UX problems. Thus, 
the implication is that 
when UX practitioners recruit participants to collect a large amount of TA usability test videos, such as via online remote usability testing (e.g., ~\cite{thompson2004here,andreasen2007happened}), to train AI models, they could consider to recruit both native and non-native English speakers. This would significantly increase their chance and lower their cost of getting a sufficient amount of TA test sessions to train AIs. 





\subsection{Limitations and Future Work}
Our research contributes to a first understanding of the effects of English language proficiency on think-aloud verbalizations and how TA verbalizations of native and non-native English speakers indicate UX problems. In this section, we would like highlight the limitations of our current work and discuss potential future work. 

First, our study included a relatively small number of participants ($N=18$). It is essential to replicate the study with more participants of diverse backgrounds to further validate and extend the findings. 

Second, we chose a subgroup of non-native English speakers---Chinese students who studied in US universities. Other non-native English speakers, such as ones from other countries, may think aloud differently in English. 
Further, our non-native English speakers had intermediate or higher language proficiency (Table~\ref{tab:participants}). Non-native English speakers with lower proficiency might verbalize their thoughts differently too. Thus, it is worth investigating how cultural backgrounds and language proficiency of non-native English speakers might affect their thinking aloud in English.

Third, we focused on English language. Other language speakers might organize their thoughts differently. Slobin argued that language ``is a subjective orientation to the world of human experience, and this orientation affects the ways in which we think while we are speaking''~\cite{slobin1996fromthought}. Indeed, research suggested that cultures and languages could affect users' think-aloud processes~\cite{hall2004cultural,clemmensen2009cultural,clemmensen2011templates}. Thus, it is worth exploring whether subtle verbalization patterns are telltale signs of UX problems for other languages, for example, Chinese, Dutch and French to name a few. If similar verbalization patterns indeed exist in other languages, then such patterns could be utilized to inform the design of AIs or AI-assisted tools to support UX evaluators to uncover UX problems for products used by people who speak those languages.

Fourth, conducting TA usability testing relies on the moderator's skills too, especially the interactive think aloud (ITA). Our study moderator had two years of UX experience in conducting TA sessions. It is possible that the moderator's experience would affect how she prompts participants (e.g., what prompts to use and how frequently to prompt participants). Thus, it is worth exploring how the moderator's experience might affect how participants think aloud.



Fifth, we selected three types of websites to increase the variations among the test products. However, it remains unknown whether and how products might affect participants' thinking aloud processes.
Similarly, task difficulty might influence how participants think aloud too. For example, participants might find it harder to verbalize their thoughts when working on a more difficult and demanding task. Thus, more research is warranted to understand how products and task difficulty might affect non-native English speakers' verbalizations and their correlations with UX problems.  

Lastly, we used three types of \textit{concurrent} think-aloud protocols in the study. To reduce the potential drawbacks of concurrent protocols, \textit{retrospective} think-aloud protocols (RTAs) are used in usability testing~\cite{fan2020Survey}. When using RTAs, participants complete the task first and then verbalize their thought processes while watching their session recording. Verbalizations in RTAs are shown to be a valid representative of participants' thoughts~\cite{guan2006validity}. It is worth investigating how participants' think-aloud verbalizations indicate UX problems in RTAs in the future.

\section{Conclusion}
Recent studies showed that subtle patterns in TA verbalizations are telltale signs of UX problems. However, such studies were conducted with native English speakers~\cite{fan2019concurrent,Fan2021OlderAdults}. There are more non-native English speakers around the world, who might think and verbalize thoughts differently than native speakers due to different cultural backgrounds. In this research, we took a first step to explore this problem space by studying a subgroup of non-native English speakers---Chinese students who study in US universities. We have compared non-native English speakers' verbalizations in three common types of concurrent TA protocols (i.e., the classic TA protocol~\cite{ericsson:1984}, the speech-communication TA protocol~\cite{boren2000thinking}, and the interactive TA protocol~\cite{hertzum2018usability,rubin2008handbook,dumas1999practical}) with those of native English speakers. Our findings show that for both non-native and native English participants, their verbalization categories and the relative proportions, the UX problems that they encountered, and the correlations between the verbalization categories and the UX problems were largely similar. Moreover, the findings were mostly consistent for three TA protocols.

Analyzing TA test session recordings to identify UX problems often entails reviewing video recordings and listening to users' think-aloud verbalizations. This process is often arduous and time-consuming, which has motivated researchers to explore ways to automate or semi-automate this analysis process~\cite{fan2020automatic,fan2020vista}. Such methods leveraged subtle patterns in users' think-aloud verbalizations and speech patterns~\cite{fan2019concurrent,Fan2021OlderAdults}. Our findings support that subtle verbalization patterns uncovered in previous studies with native English speakers~\cite{fan2019concurrent,Fan2021OlderAdults} are largely applicable to non-native English speakers. One implication is that UX practitioners could recruit both native and non-native English speakers to participate in TA usability testing to gather a larger amount of data, which could be used to train AI models to detect UX problems automatically or semi-automatically~\cite{fan2020automatic,fan2020vista,grigera2017automatic,paterno2017customizable,harms2019automated,jeong2020detecting}. As an initial exploration of this problem space, we only studied one subgroup of non-native English speakers. As culture can affect thinking and speaking behaviors, future work should investigate TA verbalizations of different subgroups of non-native English speakers and of different languages to better inform the design of automatic- or semi-automatic analysis methods for identifying UX problems that users of different cultures and languages might encounter.





\bibliographystyle{ACM-Reference-Format}
\bibliography{references.bib}

\end{document}